\documentclass[twocolumn,amssymb,showpacs,amsmath,aps,prl,superscriptaddress]{revtex4}
\usepackage{graphicx}
\usepackage{colordvi}

\begin{document}
\preprint{}

\title{ Approaching the ground state of the kagom\'e antiferromagnet}

\author{W. Schweika}
\affiliation{Institut f\"ur Festk\"orperforschung,
Forschungszentrum J\"ulich, D-52425 J\"ulich, Germany}
\author{M. Valldor} \affiliation
{Institut f\"ur Physik der Kondensierten Materie, TU-Braunschweig, D-38106 Braunschweig,
Germany}
\author{P. Lemmens}\affiliation
 {Institut f\"ur Physik der Kondensierten Materie, TU-Braunschweig, D-38106 Braunschweig, Germany}
\date{\today}

\begin{abstract}

Y$_{0.5}$Ca$_{0.5}$BaCo$_4$O$_7$ contains kagom\'e layers of Co ions, whose spins are
strongly coupled according to a Curie-Weiss temperature of -2200 K. At low temperatures,
T = 1.2 K, our diffuse neutron scattering study with polarization analysis reveals
characteristic spin correlations close to a predicted two-dimensional coplanar ground
state with staggered chirality. The absence of three dimensional long-range AF order
proves negligible coupling between the kagom\'e layers. The scattering intensities are
consistent with high spin $S=3/2$ states of Co$^{2+}$ in the kagom\'e layers and low
spin $S=0$ states for Co$^{3+}$ ions at interlayer sites. Our observations agree with
previous Monte Carlo simulations indicating a ground state of only short range chiral
order.
\end{abstract}


\maketitle


In low dimensions, antiferromagnetic (AF) order is suppressed at finite
temperatures \cite{MW66} and geometrical frustration raises the complexity of ground states
with finite entropy and non-collinear, chiral spin structures \cite{Moessner}. Spin chirality is
related to unique universality classes \cite{Kawamura,VP2006} and has relevance for the
spin excitations in  copper-oxide superconductors \cite{JT2004,Lindgard2005}.
A challenge for theoretical understanding for
decades  is the utmost frustration among AF coupled spins on the two-dimensional
kagom\'e lattice \cite{Syozi}.

Considering the configurational energy of the classical Heisenberg AF,
\begin{equation}
H= \sum_{o,r} J (\vec r) \vec S_o \cdot \vec S_r,
 \end{equation}
for  AF interaction  to only nearest neighbors, $J (|\vec r_1|) >0$,
the ground state is highly degenerate.
There are two competing ordered coplanar spin structures (see Fig.~1),
with relative spin orientations of 120 degrees, which are  either of
uniform or staggered chirality.
 We use the convention that the chirality is positive
when the spins rotate in steps of 120 degrees as one goes around a triangle.
Apart from the degeneracy of a common in-plane spin rotation,
 there are further
degeneracies related to possible local disorder, as shown in Fig.~1,
Within its configurational space, thermal fluctuations
 \cite{Huse1992,Reimers} as well as quantum fluctuations \cite{sachdev1992}
select the  $\sqrt{3} \times \sqrt{3}$ structure.
The entropical selection resolving a ground state degeneracy is a mechanism
that has been named {\em order by disorder} \cite{Villain}.
The classical ground state has still a macroscopic manifold,
which could be removed by quantum fluctuations \cite{Yildirim}.
On the other hand, there are predictions of a
disordered ground state, a quantum spin liquid with
singlet formation for the $S=1/2$ kagom\'e AF \cite{Elser}.

The topology of the classical ground state degeneracies has been analysed in more detail
in Monte Carlo (MC) simulations  of the 2D Heisenberg AF by Reimers and Berlinsky
[\onlinecite{Reimers}]. An intriguing result of these simulations is (an  indication of)
a divergence of the spin-correlation function for T $\rightarrow$ 0 K towards an ordered
ground state resembling the $\sqrt{3} \times \sqrt{3}$ structure, whereas chiral
long-range order appears to be prohibited by disorder due to the local common zero
energy spin rotations  as depicted in Fig.~1.

\begin{figure}[t]
\includegraphics[height=4.2cm]{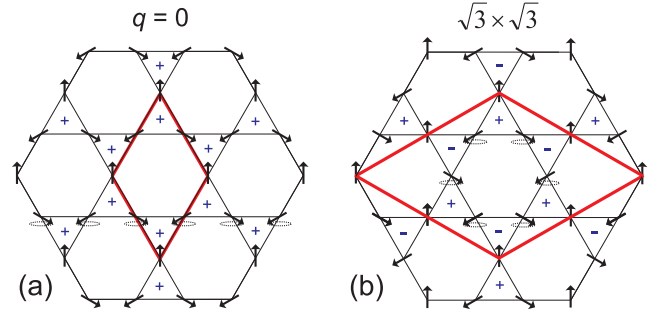}
\caption{(color online). Infinitely degenerate AF ground states on the kagom\'e lattice.
(a) the $q_0$-structure with uniform chirality; (b) the staggered chiral structure,
which has a larger unit cell by $\sqrt{3}\times  \sqrt{3}$. Chirality is denoted by
signs inside the triangles. The dashed ellipses represent thermal fluctuations of zero
energy:
 (a) common spin rotations in an infinite chain;
 (b) the {\em weathervane} defect, a
common local spin rotation on the hexagon around the axis of the spins on the
edges of the unit cell.
 }
\label{GS}
\end{figure}

Experimental studies of the 2D kagom\'e  AF are difficult and rare,
although pyrochlores \cite{Harris1997},
magnetoplumbites \cite{Broholm1990}, and jarosites \cite{Wills2001,Grohol2005,Matan2006}
contain layers of the kagom\'e structure.
For weak interlayer coupling, these candidates combine all the desired ingredients:
low dimensionality, low coordination, and strong frustration.
A neutron scattering study on magnetoplumbite SrCr$_{8-x}$Ga$_{4+x}$O$_{19}$ \cite{Broholm1990}
provided the first experimental evidence of short-range spin correlations down to 1.5 K
and indicated a preference for the $\sqrt{3} \times \sqrt{3}$  structure.
The jarosite  (D$_3$O)Fe$_3$(SO$_4$)$_2$(OD)$_6$ stays disordered even at low $T$ according to
 diffuse neutron scattering data (obtained with polarization analysis), but
 the ordering wave vector could not be clearly identified.
Another close realization of the kagom\'e AF is the jarosite
KFe$_3$(SO$_4$)$_2$(OD)$_6$, although the interference of 3D N\'eel order below 65 K
complicates comparisons to the ground state properties of the kagom\'e AF. Recent single
crystal studies show spin correlations towards the $q_0$-structure \cite{Grohol2005},
contrary to theoretical expectations for the kagom\'e AF with only nearest neighbor
interactions. Indeed, small AF interactions to next nearest neighbors, which have been
deduced from the low energy spin wave excitations \cite{Matan2006},
 stabilize the  $q_0$-structure instead of an {\em order by disorder} mechanism.
More interestingly,  the low energy spin wave excitations could be related to the
 ``zero energy modes'' as shown in Fig.~1(a) \cite{Matan2006}.

Here, we present results of diffuse neutron scattering with polarization analysis on a
new compound Y$_{0.5}$Ca$_{0.5}$BaCo$_4$O$_7$, a realization of the kagom\'e  AF with
only nearest neighbor interactions, which allows an unprecedented approach to its ground
state properties.

\begin{figure}[t]
\includegraphics*[
angle=0, scale=0.4]{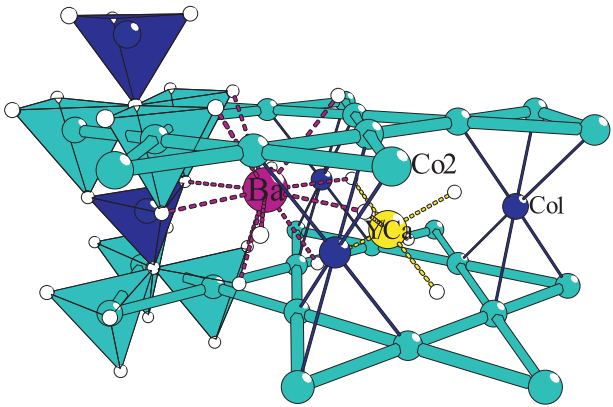} \caption{(color online). The average crystal structure of
Y$_{0.5}$Ca$_{0.5}$BaCo$_4$O$_7$. Thicker lines between the Co2 atoms highlight the
kagom\'e layers. All Co atoms are tetrahedrally coordinated by oxygen, as shown in the
polyhedral representation (left), emphasizing the three-fold symmetry axes, going
through Co1(blue), which allow trimerization of Co2 in the kagom\'e substructure. }
\label{structure}
\end{figure}
The average crystal structure of Y$_{0.5}$Ca$_{0.5}$BaCo$_4$O$_7$ is known \cite{Valldor2002,Valldor2006}
 (Fig.~2) and a mixed valence of Co$^{2+}$ and  Co$^{3+}$  is a consequence of stoichiometry.
Co occupies two different tetrahedral sites in the structure: 75\% of all Co (Co2)
constitute the kagom\'e layers and the remaining 25\% fills the Co1 sites, which are on
threefold axes between the kagom\'e layers. Within this symmetry, a trimerization in the
kagom\'e layers is possible and different Co2-Co2 bond lengths are in fact observed at
300 K \cite{Valldor2006}. Because of shorter Co-O distances, the valence of Co1 must be
higher than that of Co2, indicating a preference for   Co$^{3+}$  (3d$^6$
configuration). The bonding situation and spin state of  Co$^{3+}$  seem to be
responsible for two dimensional magnetic behavior: (i) geometrically, the tetrahedral
coordination causes a high frustration for any possible AF exchange between the kagom\'e
layers; (ii) the possibility of a low spin state of  Co$^{3+}$  at the interlayer sites,
as  Co$^{3+}$ is situated closer to a trigonal plane changing the point-group symmetry
from $T_d$ to $C_{3v}$. According to Hund's rules, an ideal tetrahedral coordination
with weak crystal field splitting (oxygen ligands) results in high spin states, on the
other hand low spin states have been observed for $C_{3v}$ symmetry
\cite{Byrne1986,Jenkins2002}.


Polycrystalline Y$_{0.5}$Ca$_{0.5}$BaCo$_4$O$_7$  samples for the diffuse neutron scattering
experiments have been synthesized by a solid-state reaction in air \cite{Valldor2006}.
 Under the applied synthesis conditions, the oxygen stoichiometry should be ideal
as verified by iodometric titration on the parent compound YBaCo$_4$O$_7$ \cite{Karppinen2006}.
The scattering  experiments were performed on the DNS instrument in J\"ulich. In
contrast to most previous investigations \cite{Harris1997,Broholm1990,Grohol2005}, the
magnetic scattering was separated from nuclear scattering by polarization analysis
\cite{Wills2001,WS_NN_2005}. {\em A fortiori}, the separated high background due to
nuclear spin-incoherent scattering from Co serves as an intrinsic calibration of the
paramagnetic cross-section; a procedure that avoids usual systematic errors and
corrections.
\begin{figure}[h]
\includegraphics*[
angle=0, scale=0.525]{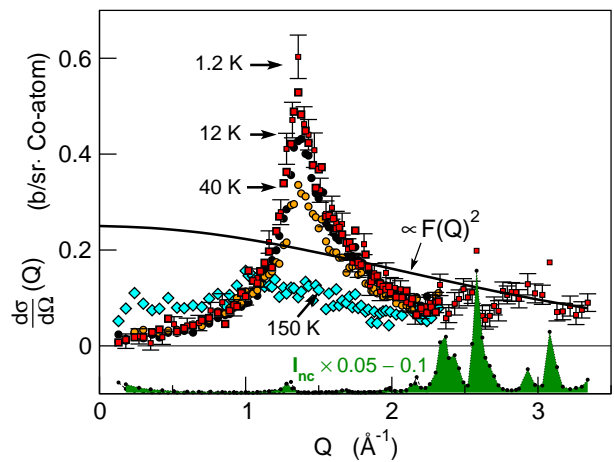} \caption{(color online). Separation of scattering
contributions and temperature dependence of the separated diffuse magnetic scattering;
data are calibrated by the separated spin-incoherent scattering of Co (0.382 b/sr);
 the separated nuclear coherent scattering, I$_{nc}$ has been
rescaled and shifted  (bottom part).
}
\label{DNS-data}
\end{figure}
A strong diffuse peak evolves upon cooling and its asymmetric shape indicates low dimensional
spin correlations (Fig.~3). Remarkably, no long-range magnetic order appears even at very low T.
The magnetic scattering can be described in terms of the Fourier transform of the
spin pair-correlations,
applying orientational averaging for the polycrystalline sample \cite{Blech}.
For a direct comparison with the MC results, we use a simplified scattering expression \cite{Reimers},
which neglects a possible interrelationship between spin correlation and spin direction:
\begin{eqnarray}
\frac{d\sigma }{ d \Omega}_{mag} & =&   \frac{2}{3}  \left( \frac{e^2 \gamma}{m_e c^2}\right)^2  F_Q^2
\sum_r   \langle \vec S_0 \cdot \vec S_r\rangle  \frac{\sin Qr}{Qr }.
\end{eqnarray}
Here, the product of the classical electron radius $(e^2/m_ec^2) = 0.282 \times 10^{-12} \mathrm{cm}$
and the neutron gyromagnetic ratio = -1.91 defines the magnetic scattering length
$r_0 = -0.54 \times  10^{-12} \mathrm{cm}$; $m_e$ is the electron mass, $e$ is its charge,
and $c$ is the velocity of light. $S$ represents the spin quantum number of the scattering ion,
$F_Q$ denotes the magnetic scattering form-factor of the single ion Co$^{2+}$.
For the ideal paramagnet, $T \gg J$, the scattering is purely elastic and proportional to the
self-correlation $S(S+1)$.
In general, one has to consider the expectation value of the time-dependent self-correlation function
of a spin precessing in the strong local field of its neighbors.
For $T \ll J$ and integrating over only small energy transfers in diffraction,
the expectation value reduces effectively to that of the scalar product of the
ordered moments $\langle S^2_z \rangle$.

%
The magnetic scattering intensity allows us to estimate the average Co spin moment.
A fit to the 1.2 K data yields a forward scattering of 0.25(2) b/sr$\cdot$Co for the self-correlation term.
The best quantitative agreement is found if we assume that Co$^{2+}$ ions (5/8 of all Co) are in
high spin state $S = 3/2$ and that Co$^{3+}$ ions (3/8 of all Co)
 are in low spin state $S = 0$ yielding 0.273 b/sr$\cdot$Co-ion.
It is clear that $S = 0$ at the Co1 sites rationalizes the absence of any
significant interlayer coupling and, hence, the ultimate suppression
 of 3D long-range order.

The diffuse magnetic scattering
probes the wave-vector dependent susceptibility
$\chi(Q) = \sum_r \langle \hat S_0 \cdot \hat S_r \rangle {\sin Qr \over Qr}$,
where  $\hat S$ are spins of unit length.
Within the approximation made for Eq.~(2), the 1.2 K data are compared with the MC results
in Fig.~4.
%
%
 Correlations between the directions of the spins and their distance vector
 would require  higher order corrections neglected in Eq.~(2).
In case of the   $\sqrt{3} \times \sqrt{3}$  structure, as shown in Fig.~1(b), we
analyzed that this would result in  significant intensities for $0 < Qr < 1$, for
nearest neighbor distances $|\vec r_1| = a/2$ and a unit cell parameter $a =6.30 \,
\mathrm{\AA}$, which is in contradiction to our observations.
According to Eq.~(1), the Hamiltonian 
  is degenerate with respect to  a common  spin rotation in the kagom\'e plane,
and we conclude that this degeneracy  is essentially preserved.
%
%

Furthermore, we can rule out spin correlations towards the $q_0$-structure,
whose ordering wave vectors are  $a^*(h,k)$;  the strongest peak would be at
$Q=|\mathbf{q}_0| = |a^*(1,0)|=1.152\;\mathrm \AA^{-1}$, where $a^* = (2\pi/a) (2/\sqrt{3})$.
Instead of this,  the observed peak coincides with
$Q=|2 \mathbf{q}_{\sqrt{3}}| = |2a^*(1/3,1/3)|=1.330 \;\mathrm \AA^{-1}$, the ordering wave vector of
the $\sqrt{3} \times \sqrt{3}$  structure.

\begin{figure}[h]
\includegraphics*[
angle=0, scale=0.525]{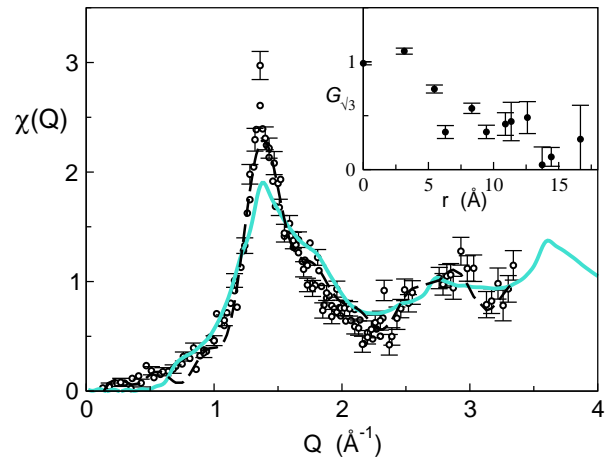} \caption{(color online). Susceptibility $\chi(Q)$ from
1.2 K data  {\em versus} modulus of wave vector Q;
the thick (cyan) line represents MC data from Ref.~\cite{Reimers}.
for a classical Heisenberg AF at T/J = 0.002;
dashed line represents a fit of the spin correlation by Fourier analysis,
whose results, as normalized to correlations in the ideal  structure,
$G_{\sqrt{3}}$,  are shown in the inset.
}
\label{DNS-model}
\end{figure}

%
The data has been analyzed in terms of spin correlations by a
 Fourier analysis using a linear least squares refinement.
The results are shown in the inset of Fig.~\ref{DNS-model}, where the
spin correlations have been normalized to the correlation function
of the ideal  $\sqrt{3} \times \sqrt{3}$ structure:
\begin{equation}
  G_{\sqrt3}(r) =\langle  \hat  S_0 \cdot \hat S_r \rangle
                   / \cos \mathbf q_{\sqrt3} \cdot \mathbf r,
\end{equation}
which equals to unity when the spins are in the  $\sqrt{3} \times \sqrt{3}$
configuration. A limited number of parameters could be determined,
which clearly show the preference for building up the $\sqrt{3} \times \sqrt{3}$
structure.

%
%

An interesting question is whether our data can provide any indication for
the proposed {\em weathervane} defects (cf. Fig.~1(b)) that seem to disturb the chiral
long range order of the ground state in the previous MC simulations.
Therefore, one may compare a model of the ideal  $\sqrt{3} \times \sqrt{3}$
 structure to a defect model, and convolute the correlations with an adjustable
asymptotic decay function.
A simple defect model is to place the {\em weathervane} defects on two of the three sublattices.
Note that the configurational average of this model is equivalent to a collinear structure model,
and that the correlations are reduced by a factor 2 with respect the ideal structure
( $ \langle \hat S_0 \cdot \hat S_r  \rangle = 1$  or  $-1/2$)
 for distances exceeding the spacings that occur within the unit cell.
An algebraic decay of the correlations 
has been found in MC simulations \cite{Reimers}, which relates the spin correlations to
a percolation phenomenon. However,  the present data are affected by instrumental
resolution (see Fig.~3  separated nuclear scattering)
 such that we cannot reliably distinguish between an exponential and algebraic decay.
Models based either on the ideal structure or on the {\em weathervane} defect
 are in close agreement with the data. The exception is that both models produce
a small additional peak at $q_{\sqrt{3}}$, an integral intensity that is one order of
magnitude weaker than the peak at  $2q_{\sqrt{3}}$. The discrepancy is less significant,
approximately half the size, if the {\em weathervane} defects are included. Apparently,
the spin structure is even less ordered and a more appropriate model would include that
{\em weathervane} defects are distributed on all sublattices.
%
%
Therefore, MC simulations are adequate to sample the  configurational space. Indeed,
there is a  qualitatively astonishing agreement between our data and the MC results
\cite{Reimers} for the classical Heisenberg AF. In particular, MC simulations reproduces
the absence of the peak at  $q_{\sqrt{3}}$. The disorder has been related to a structure
of chiral domains with anti-phase boundaries. According to the MC data, the correlation
length of chiral order parameter stays short ranged for $T \rightarrow 0 \; \mathrm K$ ,
while the spin correlation length and the peak at  $2q_{\sqrt{3}}$ diverge in this
limit.
It is noteworthy that this observed peak, 
as compared to the MC simulation,
 is significantly more pronounced. Hence, the ground state is
approached much closer in our experiment,
although temperature and Curie-Weiss constant of approximately $- 2200\, \mathrm K
\approx 4J$ \cite{Valldor2006}
 matches closely the conditions of the simulation,  $T/J = 0.002$.
In fact, the peak is affected by the instrumental resolution
and the true peak height is already substantially reduced. 

The {\em weathervane} defects (see Fig.~1(b)) are local collective spin excitations
of zero energy according to Eq.~(1) that
preserve the sum of spins and give rise to residual entropy in analogy to
ice \cite{Pauling} and spin-ice systems \cite{Harris1997},
Local collective spin excitations have also been observed in
inelastic neutron scattering from cubic spinel ZnCr$_2$O$_{4}$ \cite{Lee2002}.
 Very recently, ``chain-like'' excitations (from the $q_0$-structure, see Fig.~1(a))
are lifted to finite energies by Dzyaloshinskii-Moriya (DM) interaction
as been reported from inelastic neutron scattering \cite{Matan2006}.
We performed additional time-of-flight experiments verifying that the scattering
at 1.2 K is essentially elastic, and proving that the spin correlations are frozen
within the energy resolution of 0.5 meV.
At higher temperatures, the  inelastic scattering  appears gapless, indicating the absence of
anisotropies related to DM interaction, and exhibits a quasi-elastic broadening similar
to the observations on SrCr$_{8-x}$Ga$_{4+x}$O$_{19}$ \cite{Broholm1990}.

The strong AF exchange in the plane causes a dynamic response at high energies,
which suggests that the low-energetic relaxations correspond to
thermal spin fluctuations out of the kagom\'e plane.
Theoretically, the complex free energy landscape of the kagom\'e AF will induce very
slow spin dynamics similar to ordinary spin glasses \cite{Ferrero2003} and may explain
observed field cooling effects at 370~K \cite{Valldor2006}.

To conclude, the present neutron study with polarization analysis on a new compound,
realizing the $S=3/2$ kagom\'e AF, agrees in surprising detail with the complexity of
the classical ground state for only nearest neighbor interactions
as has been found in previous MC simulations. 
 The remarkable survival of only short-range 2D spin
correlations at low temperatures is due to severe geometrical frustration by tetrahedral
coordination combined with high/low spin states of Co ions in alternating layers. Site
disorder with spin dilution, albeit beyond the site percolation threshold \cite{Sykes},
results from assuming ionic bonding and from exchange of moments through electron
hopping. Covalency effects might restore the predicted behavior in the experiment, which
implies a long-range ordered ground state with only short-range chiral order.

We thank J.N. Reimers and A.J. Berlinsky for providing us with their original MC data.


\end{document}